\documentclass[a4paper,11pt]{article}
\usepackage{pos}

\usepackage{slashed}

\numberwithin{equation}{section}
\usepackage{tikz}
\usetikzlibrary{intersections}

\newcommand{\dd}{\mathrm{d}}

\title{Novel Lattice Formulation of 2D Chiral Gauge Theory via Bosonization}

\author[a]{Okuto Morikawa}
\author*[b]{Soma Onoda}
\author[b]{Hiroshi Suzuki}

\affiliation[a]{Interdisciplinary Theoretical and Mathematical Sciences Program (iTHEMS),
RIKEN, Wako 351-0198, Japan}
\affiliation[b]{Department of Physics, Kyushu University,
744 Motooka, Nishi-ku, Fukuoka 819-0395, Japan}

\emailAdd{okuto.morikawa@riken.jp}
\emailAdd{onoda.soma@phys.kyushu-u.ac.jp}
\emailAdd{hsuzuki@phys.kyushu-u.ac.jp}

\abstract{Recently, lattice formulations of 2D Abelian chiral gauge theory
have been constructed based on Abelian bosonization.
It is remarkable about these 2D lattice formulations that
they reproduce the same gauge anomaly structure as the continuum theory,
even at a finite lattice spacing.
In this talk, we propose yet another lattice formulation
based on the ``excision method'' introduced recently in Ref.~\cite{Abe:2023uan}.
This approach respects the admissibility condition,
which is a constraint on the smoothness of lattice field configurations;
it usually prohibits magnetically charged objects,
that is, vector-charged objects in fermion theories.
We show that such objects can be defined in the excision method as a lattice defect called a ``hole,'' and discuss the selection rules for charged objects.}

\FullConference{The 41st International Symposium on Lattice Field Theory (LATTICE2024)\\
 28 July - 3 August 2024\\
Liverpool, UK\\}

\begin{document}
\maketitle

\section{Introduction}
Chiral gauge theory is the foundation for many significant physical theories,
such as the Standard Model and Grand Unified Theories.
However, a non-perturbative definition of chiral gauge theory
remains a considerable challenge for a long time.
For the suitable matter content that
the gauge anomaly cancellation condition is met in the continuum,
how is the cancellation mechanism reproduced in a non-perturbative construction,
e.g., a lattice gauge theory?

There have been many studies on four-dimensional
lattice chiral gauge formulations; we succeeded in getting
the $U(1)$ case~\cite{Luscher:1998du} and
the $SU(2)\times U(1)$ theory~\cite{Kikukawa:2000kd,Kadoh:2007xb}.
Recently, the authors of Refs.~\cite{DeMarco:2023hoh,Berkowitz:2023pnz} focused
on lattice formulations of two-dimensional~(2D) chiral gauge theory
based on Abelian bosonization~\cite{Coleman:1974bu,Mandelstam:1975hb}.
Bosonization in 2D, which utilizes (chiral) compact bosons
instead of chiral fermions,
enables the construction of a lattice theory that belongs
to the same universality class as the desired chiral gauge theory.
The good news is that, in this construction,
the gauge anomaly structure can be straightforwardly derived
at the \textit{classical level},
which is identical to that in the continuum theory.
Consequently, exact chiral gauge symmetry can be realized
even at a finite lattice spacing.

Compared to the previous works, we propose a ``novel lattice formulation''
with dual vertex operator, $e^{i\Tilde{\phi}}$. (See Ref.~\cite{Morikawa:2024zyd}.)
This operator corresponds to a magnetic object
in the context of compact scalar theory, and is also the bosonic counterpart
to vector-symmetry charged objects in chiral fermion theories.
Especially, our formulation respects
the admissibility condition~\cite{Luscher:1981zq,Luscher:1998du,Hernandez:1998et},
which imposes a smoothness condition
on lattice fields for each lattice site~$n$.
To define a vector-charged object of field configurations
under the admissibility condition,
we represent a magnetic object in a compact boson system as a ``hole,''
a lattice defect region excised from the lattice spacetime;
now this hole is supposed to be located at the dual lattice, $\Tilde{n}$.
This approach is based on the excision method proposed in Ref.~\cite{Abe:2023uan}.
In general, a non-zero winding number of compact boson~$\phi$ around the hole
is compatible with the admissibility condition.

The organization of this paper is as follows.
In Section~\ref{sec:continuum}, we discuss
the 2D $U(1)$ chiral gauge theory in the continuum,
on which our lattice formulation is based.
In Section~\ref{sec:lattice}, we describe
the lattice formulation of the theory introduced in Section~\ref{sec:continuum}.
First, we introduce the compact boson defined on the lattice
and the $U(1)$ gauge field.
We then define vector-charged objects using the excision method,
which is a unique point of this talk.
Next, we define a lattice action based on bosonization
and demonstrate that the gauge anomaly structure of the continuum theory
can be easily reproduced at the classical level
with exact chiral gauge symmetry even at a finite lattice spacing.
Finally, in Section~\ref{sec:selection}, we show that
the selection rules for charged objects defined in Section~\ref{sec:lattice}
are consistent with the fermion number anomaly in the continuum theory.

\section{2D $U(1)$ chiral gauge theory:
Continuum formulation and bosonization}\label{sec:continuum}

The fermion action for the chiral gauge theory on a 2D manifold~$M_2$
of interest is given by
\begin{align}
    S_F &:= \int_{M_2} \dd^2x \, \sum_{\alpha} \Bar{\psi}_\alpha
    \left[i\slashed{\partial} - \slashed{A}(q_{R,\alpha}P_R + q_{L,\alpha}P_L)\right]
    \psi_\alpha\\
    &= \int_{M_2} \dd^2x \, \sum_{\alpha} \Bar{\psi}_\alpha
    \left[i\slashed{\partial} - \slashed{A}(q_{V,\alpha} + \gamma_3 q_{A,\alpha})\right]
    \psi_\alpha,
    \label{eq:fermion-continuum}
\end{align}
Here, $\alpha = 1$, \dots, $N_f$ denotes flavor degrees of freedom,
$A_\mu$ is the $U(1)$ gauge potential, and
\begin{align}
    P_{R,L} &:= \frac{1\pm\gamma_3}{2},\quad
    q_{V,\alpha} = \frac{q_{R,\alpha} + q_{L,\alpha}}{2}, \quad q_{A,\alpha} = \frac{q_{R,\alpha} - q_{L,\alpha}}{2}.
\end{align}
Note that each left- and right-handed fermion is gauged
with an independent charge assignment, $q$.
Gauge anomaly generally arises as
\begin{align}
    \text{(gauge anomaly)} = -\frac{i}{2\pi}
    \left(\sum_\alpha q_{A,\alpha}q_{V,\alpha}\right)
    \int_{M_2} \dd^2x \, \Lambda(x) F_{12},
\end{align}
where $\Lambda(x)$ is the gauge transformation parameter,
and $F_{12}$ is the field strength.
Therefore, the anomaly cancellation condition is
\begin{align}
    \sum_\alpha q_{A,\alpha}q_{V,\alpha}
    = \frac{1}{4}\left(\sum_\alpha q_{R,\alpha}^2 - q_{L,\alpha}^2\right) = 0.
\label{eq:anomaly-cancel-continuum}
\end{align}

One finds the bosonic counterpart of $S_F$,
which can be constructed to share the same gauge anomaly structure based on bosonization,
\begin{align}
\int_{M_2} \dd^2x \, \sum_{\alpha}
\left[
   \frac{R^2}{4\pi}\sum_{\mu}(\partial_\mu\phi_{\alpha} + 2q_{A,\alpha}A_\mu)^2 + \sum_{\mu,\nu}\frac{iq_{V,\alpha}}{2\pi}A_\mu\epsilon_{\mu\nu}(\partial_\nu\phi_{\alpha} + 2q_{A,\alpha}A_\nu) 
\right] ,
\label{eq:continuum-bosonic-action}
\end{align}
where, $\phi_\alpha$ is a compact boson with a $2\pi$ periodicity
($\phi_\alpha \sim \phi_\alpha + 2\pi$),
and the parameter $R$ is the compactification radius.\footnote{In this talk, we set $R^2 = 1/2$ to represent the 2D fermion.}

According to the bosonization rules (see, e.g., \S7.5 of Ref.~\cite{Tong:18}),
axial symmetry corresponds
to the electric symmetry (shift symmetry) of $\phi_\alpha$,
while vector symmetry corresponds
to its magnetic symmetry (winding symmetry).
To be consistent with the compactness $\phi_\alpha \sim \phi_\alpha + 2\pi$, $2q_{A,\alpha}$ and $q_{V,\alpha}$ must be integers. 
Then, we note that models like the so-called $21111$ model,
where $q_{V,\alpha}$ can be a half-integer,
fall outside the scope of the present bosonization-based approach.

\section{Lattice formulation of 2D $U(1)$ chiral gauge theory}\label{sec:lattice}
\subsection{Lattice field contents and definitions}
To construct a lattice counterpart of the chiral gauge theory
represented by the action~\eqref{eq:continuum-bosonic-action},
we consider a compact boson theory on $M_2=T^2$
approximated by a square lattice.
Respecting the $2\pi$ periodicity,
we define the following dynamical variables at each site $n$ on the lattice:
\begin{align}
    e^{i\phi_\alpha(n)},\quad -\pi\leq\phi_\alpha(n)<\pi.
\end{align}
We also introduce the $U(1)$ gauge potential as link variables and define the field strength as a logarithm of the so-called plaquette term:
\begin{align}
U_\mu(n)&=e^{iA_\mu(n)},& -\pi&\leq A_\mu(n)<\pi,\quad \mu =1,2,\\
F_{\mu\nu} (n)&:=\frac{1}{i} \ln [U_{\mu} (n)U_{\nu} (n+\Hat{\mu})U_{\mu} (n+\Hat{\nu})^{-1}U_{\nu}(n)^{-1}]\notag\\
      &= \Delta_\mu A_{\nu} (n) -\Delta_\nu A_{\mu} (n) +2\pi N_{\mu\nu}(n),& -\pi&\leq  F_{\mu\nu} (n) <\pi.\label{eq:field-strength}
\end{align}
Here, $\Delta_\mu$ denotes the forward difference,
$\Delta_\mu f (n):=f (n+\Hat{\mu})-f(n)$.
A branch of the logarithm adopts the principal branch,
and $N_{\mu\nu}(n)$ is an integer ensuring that $F_{\mu\nu}(n)$ resides in this branch.

For technical reasons, we introduce the dual lattice $\Tilde{M}_2$, where $\Tilde{n}:=n+\frac{1}{2}(\Hat{1}+\Hat{2})$, and place a copy of the link variable
\begin{align}
U_\mu(n)=U_\mu(\Tilde{n}).
\label{eq:copy-of-link}
\end{align}
As the lattice counterpart of the ``covariant derivative''
$\partial_\mu\phi_{\alpha}+2q_{A,\alpha}A_\mu$ in the continuum theory,
we define the following quantity:
\begin{align}
    D\phi_\alpha(n,\mu)&:= \frac{1}{i} \ln \left[ e^{-i\phi_\alpha(n)} U(n,\mu)^{2q_{A,\alpha}} e^{i\phi_\alpha(n+\Hat{\mu})} \right]\notag\\
&=\Delta_\mu\phi_\alpha(n)+2q_{A,\alpha}A_\mu(n)
   +2\pi\ell_{\alpha,\mu}(n),\label{eq:covariant-derivative}
\end{align}
where $\ell_{\alpha,\mu}(n)$ is an integer that ensures the principal branch, similar to $N_{\mu\nu}(n)$.  

Gauge transformations are defined as follows:
\begin{align}
e^{i\phi_\alpha(n)}&\to e^{i\phi_\alpha(n)}e^{-2q_{A,\alpha}i\Lambda(n)},\notag\\
   U(n,\mu)&\to e^{-i\Lambda(n)}U(n,\mu)e^{i\Lambda(n+\Hat{\mu})},\notag\\
   U(\Tilde{n},\mu)&\to e^{-i\Lambda(\Tilde{n})}U(\Tilde{n},\mu)
e^{i\Lambda(\Tilde{n}+\Hat{\mu})}.\label{eq:gauge-trans}
\end{align}
Respecting Eq.~\eqref{eq:copy-of-link}, we assume $\Lambda(n)=\Lambda(\Tilde{n})$. We can also express the gauge transformation~\eqref{eq:gauge-trans} as follows:
\begin{align}
\phi_\alpha(n)&\to\phi_\alpha(n)-2q_{A,\alpha}\Lambda(n),&
\ell_{\alpha,\mu}(n)&\to\ell_{\alpha,\mu}(n)-2q_{A,\alpha}L_\mu(n),\notag\\
A_\mu(n)&\to A_\mu(n)+\Delta_\mu\Lambda(n)+2\pi L_\mu(n),&
A_\mu(\Tilde{n})&\to A_\mu(\Tilde{n})+\Delta_\mu\Lambda(\Tilde{n})+2\pi L_\mu(\Tilde{n}),\notag\\
N_{\mu\nu}(n)&\to N_{\mu\nu}(n)-\Delta_\mu L_{\nu} (n) +\Delta_\nu L_{\mu} (n),&
N_{\mu\nu}(\Tilde{n})&\to N_{\mu\nu}(\Tilde{n})-\Delta_\mu L_{\nu} (\Tilde{n}) +\Delta_\nu L_{\mu} (\Tilde{n}).
\end{align}
Here, $L_\mu(n)$ is an integer defined to ensure the principal branch of $A_\mu=-i\ln U_\mu$ under gauge transformations.

\subsection{Admissibility condition and the excision method for vector charged object}

To introduce topological natures like magnetic symmetry into a lattice system, we restrict the configuration of lattice fields to be sufficiently smooth. This is achieved using a gauge-invariant condition called the admissibility condition.
We find that such a condition is defined by
\begin{align}
\sup_{n,\mu}\left|D\phi_\alpha(n,\mu)\right|<\epsilon,\quad \sup_{n,\mu,\nu}\left|2q_{A,\alpha}F_{\mu\nu}(n)\right|<\delta,\quad
\sup_{n,\mu,\nu}\left|q_{V,\alpha}F_{\mu\nu}(\Tilde{n})\right|<\delta,\notag\\
0<\epsilon<\frac{\pi}{2},\quad 0<\delta<\min(\pi, 2\pi-4\epsilon).
\end{align}
Under this condition, the following inequality holds:  
\begin{align}
\left|\Delta_\mu\ell_{\alpha,\nu}(n)-\Delta_\nu\ell_{\alpha,\mu}(n)-N_{\mu\nu}(n)\right|
&=\frac{1}{2\pi}
\left|\Delta_\mu D\phi_\alpha(n,\nu)-\Delta_\nu D\phi_\alpha(n,\mu)-F_{\mu\nu}(n)\right|
\notag\\
&<\frac{2}{\pi}\epsilon+\frac{1}{2\pi}\delta<1.
\label{eq:bound-integer}
\end{align}
This implies the identity as
\begin{align}
\Delta_\mu D\phi_\alpha(n,\nu)-\Delta_\nu D\phi_\alpha(n,\mu)=F_{\mu\nu}(n).
\label{eq:gauge-inv-Bianchi}
\end{align}

If we set $ U_\mu(n)=1 $, we recover the lattice version of the Bianchi identity $ \dd(\dd\phi_\alpha)=0 $ in the continuum theory. The admissibility condition provides a criterion for realizing magnetic (or vector) symmetry in the lattice theory. A question then arises: how should vector-charged objects be defined?

\begin{figure}[htbp]
\centering
\begin{tikzpicture}[scale=0.55]
  \draw[style=help lines,densely dashed,shift={(0.5,0.5)}] (-0.99,-0.99) grid[step=1] (5.99,5.99);
  \foreach \x in {0.5,1.5,2.5,3.5,4.5,5.5} {
    \foreach \y in {0.5,1.5,2.5,3.5,4.5,5.5} {
      \node[draw,circle,style=help lines,inner sep=0.5pt,fill] at (\x,\y) {};
    }
  }
  \draw (-0.5,-0.5) grid[step=1] (6.5,6.5);
  \fill[black!0] (2,1) -- (3,1) -- (3,2) -- (5,2) -- (5,4) -- (4,4) -- (4,5) -- (2,5) -- (2,1);
  \foreach \x / \y in {3.5/2, 5/3.5, 3.5/5, 2/3.5} {
    \draw[style=help lines,densely dashed] (\x,\y) -- (3.5,3.5);
  }
  \foreach \x / \y / \ax / \ay / \bx / \by in {3/1.5/2.7/1.5/2.9/3, 2.5/1/2.5/2.2/2.8/3, 2/1.5/2.2/1.5/2.5/3.1, 4.5/2/4.5/2.3/3.5/3.4, 5/2.5/4.7/2.5/3.6/3.5, 4.5/4/4.5/3.7/3.6/3.5, 4/4.5/3.7/4.5/3.5/3.6, 2.5/5/2.5/4.7/3.5/3.6, 2/4.5/2.3/4.5/3.4/3.5, 2/2.5/2.1/2.5/2.4/3.2,2/1.5/2.2/1.5/2.5/3.1
} {
    \draw[style=help lines,thin,densely dashed] (\x,\y) .. controls (\ax,\ay) and (\bx,\by) .. (3.5,3.5);
  }
  \draw[ultra thick] (2,1) -- (3,1) -- (3,2) -- (5,2) -- (5,4) -- (4,4) -- (4,5) -- (2,5) -- (2,4)
-- (2,1);
  \fill (3.5,3.5) circle(0.1) node[below] {$\Tilde{n}_{\ast}$};
  \draw (2.8,4.2) node {$\mathcal{D}$};
\end{tikzpicture}
\caption{A ``hole'' $\mathcal{D}$ excised from $M_2$, with the corresponding dual lattice (dashed lines). Inside the excised region, a site $\Tilde{n}_*$ of the dual lattice is located, as shown in the figure.}
\label{fig:1}
\end{figure}
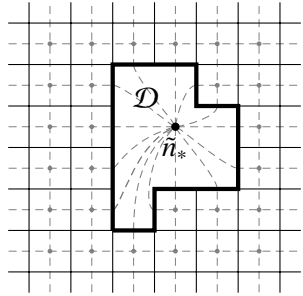

We define charged objects under vector symmetry
by excising the lattice to create a ``hole.''
This approach was proposed in Ref.~\cite{Abe:2023uan}
to introduce magnetic objects in lattice compact scalar theory (see Fig.~\ref{fig:1}).
Based on bosonization methods,
we interpret magnetic objects as vector-charged objects in the present talk.
Here, the gauge-invariant vector charge inside a loop $C$ is defined as follows:
\begin{align}
m_\alpha&:=\frac{1}{2\pi}\Bigg[\sum_{(n,\mu)\in C}D\phi_\alpha(n,\mu)-2q_{A,\alpha}F(C)\Bigg],
\label{eq:gauge-inv-charge}\\
F(C)&:=\frac{1}{i}\ln\prod_{(n,\mu)\in C} U(n,\mu).
\end{align}
The loop $C$ is topological due to Eq.~\eqref{eq:gauge-inv-Bianchi}. If there is a sufficiently large ``hole'' $\mathcal{D}$ within the loop $C$ shown in Fig.~\ref{fig:1}, $m_\alpha$ can take non-zero values. Similar to $F_{\mu\nu}(n)$, we impose a bound on $F(C)$:
\begin{align}
|2q_{A,\alpha}F(\partial\mathcal{D})|<\delta'.
\end{align}
We do not assume a specific restriction on $\delta'$ at this stage. Under these conditions, and by similar reasoning to Eq.~\eqref{eq:bound-integer}, if the number of links forming $\partial\mathcal{D}$ exceeds $2\pi/\epsilon$, $m_\alpha$ can take non-zero values.\footnote{If the number of links forming $\partial\mathcal{D}$ is comparable to that of a single plaquette, $m_\alpha$ will be zero due to the same reasoning as Eq.~\eqref{eq:bound-integer}.} This means that the ``hole'' $\mathcal{D}$ can be used as the lattice counterpart of a vector-charged object $e^{im_\alpha\Tilde{\phi}}$ in the continuum theory.\footnote{Strictly speaking, simply creating a hole does not fix $m_\alpha$, as its value may dynamically change. To define an operator with the same effect as $e^{im_\alpha\Tilde{\phi}}$, boundary conditions around $\partial\mathcal{D}$ must be fixed to yield specific $m_\alpha$.}

We note that near the ``hole,'' the one-to-one correspondence between the links of the original lattice and those of the dual lattice does not hold in general. Thus, in such cases, the relation~\eqref{eq:copy-of-link} breaks. In such cases, we treat the ``extra'' link variables as independent degrees of freedom.

\subsection{Lattice action and gauge anomaly}

The lattice action corresponding to Eq.~\eqref{eq:continuum-bosonic-action} is defined by\footnote{For the continuum limit, the radius $R^2$ should be tuned to a specific value, not the classical value $1/2$ (see, e.g., Ref.~\cite{Janke:1993va} for this issue).}
\begin{align}
S_{\mathrm{B}}&=\sum_{\alpha}\sum_{n\in M_2}\Biggl[
   \frac{R^2}{4\pi}
   \sum_{\mu}D\phi_\alpha(n,\mu) D\phi_\alpha(n,\mu)
   +\sum_{\mu,\nu}\frac{i}{2\pi}q_{V,\alpha}
\epsilon_{\mu\nu}A_\mu(\Tilde{n})D\phi_\alpha(n+\Hat{\mu},\nu)
\notag\\
   &\qquad\qquad\qquad{}
   +\sum_{\mu,\nu}\frac{i}{2}q_{V,\alpha}
\epsilon_{\mu\nu}N_{\mu\nu}(\Tilde{n})\phi_\alpha(n+\Hat{\mu}+\Hat{\nu})
   \Biggr].
\end{align}

In compact scalar theory, a mixed 't~Hooft anomaly exists between electric symmetry and magnetic symmetry, thus in general charge assignments, $ S_{\mathrm{B}} $ cannot remain invariant under the gauge transformation~\eqref{eq:gauge-trans}. This 't~Hooft anomaly corresponds to the desired gauge anomaly. Calculating the shift of~$S_{\mathrm{B}}$ under the gauge transformation~\eqref{eq:gauge-trans} yields
\begin{align}
  S_{\mathrm{B}} &\to S_{\mathrm{B}}+i\left(\sum_\alpha q_{V,\alpha}q_{A,\alpha}\right)
   \sum_{n\in M_2}\sum_{\mu,\nu}\epsilon_{\mu\nu}
\notag\\
   &\qquad\qquad{}
   \times\biggl\{
   -\frac{1}{2\pi}\Lambda(\Tilde{n})F_{\mu\nu}(n)
   -
   \left[N_{\mu\nu}(\Tilde{n})
   -\Delta_\mu L_\nu(\Tilde{n})+\Delta_\nu L_\mu(\Tilde{n})\right]
   \Lambda(n+\Hat{\mu}+\Hat{\nu})
\notag\\
   &\qquad\qquad\qquad{}+2L_\mu(\Tilde{n})A_\nu(n+\Hat{\mu})
   \biggr\}.
\label{eq:gauge-anomaly}
\end{align}

In the present bosonization-based approach, such a gauge anomaly can be computed at the classical level. Furthermore, the anomaly cancellation condition can be read from the coefficient, that is,
\begin{align}
    \sum_\alpha q_{V,\alpha}q_{A,\alpha}=0.
    \label{eq:anomaly-cancel-lattice}
\end{align}
This has exactly the same form as Eq.~\eqref{eq:anomaly-cancel-continuum} in the continuum theory. Under the condition~\eqref{eq:anomaly-cancel-lattice}, the $ U(1) $ gauge potential $A_\mu$ can be promoted to a dynamical field. Then, we can obtain a chiral gauge theory.\footnote{More precisely, a lattice theory belonging to the same universality class as Eq.~(\ref{eq:fermion-continuum}) can be constructed.}

Moreover, under the condition~\eqref{eq:anomaly-cancel-lattice}, considering the case where a vector-charged object with charge $ m_\alpha $ is inserted, we obtain the shift in $ S_{\mathrm{B}} $,
\begin{align}
   S_{\mathrm{B}} &\to S_{\mathrm{B}}-i\sum_\alpha q_{V,\alpha}
   \Lambda(\Tilde{n}_*)
   \sum_{(n,\mu)\in\partial\mathcal{D}}
   \left[
   \ell_{\alpha,\mu}(n)
   +\frac{2q_{A,\alpha}}{2\pi}A_\mu(n)
   \right]
\notag\\
  &=S_{\mathrm{B}}-i\sum_\alpha q_{V,\alpha}m_\alpha\Lambda(\Tilde{n}_*).
\label{eq:vector-trans-for-charge}
\end{align}
This shift can be interpreted as a representation of the vector gauge transformation of the vector-charged object.

\section{Selection rule}\label{sec:selection}
Let the vector charged object defined by the excision method be denoted as $M_{\{m_\alpha\}}(\mathcal{D})$. Since it shows gauge transformation properties as in Eq.~\eqref{eq:vector-trans-for-charge}, we need to attach an open Wilson line to make it gauge-invariant:
\begin{align}
    M_{\{m_\alpha\}}(\mathcal{D}) \exp\left[
   -i\sum_\alpha q_{V,\alpha}m_\alpha\sum_{(\Tilde{n},\mu)\in\Tilde{P}}^{\Tilde{n}_*}A_\mu(\Tilde{n})
   \right].
\end{align}
Here, $\Tilde{P}$ is the path with the endpoint $\Tilde{n}$. Now, for such a vector-charged object, we can derive the following selection rule from the definition of $m_\alpha$:
\begin{align}
    \sum_{\Tilde{I}}m_{\Tilde{I},\alpha}
   &=-\frac{2q_{A,\alpha}}{2\pi}\left[\sum_{p\in M_2-\sum_{\Tilde{I}}\mathcal{D}_{\Tilde{I}}}
   F_{12}(p)+\sum_{\Tilde{I}}F(\partial\mathcal{D}_{\Tilde{I}})\right]
   =-2q_{A,\alpha}Q\label{eq:vector-selection}\\
   Q&:=\frac{1}{2\pi}\left[\sum_{p\in M_2-\sum_{\Tilde{I}}\mathcal{D}_{\Tilde{I}}}
   F_{12}(p)+\sum_{\Tilde{I}}F(\partial\mathcal{D}_{\Tilde{I}})\right]\in\mathbb{Z}.\label{eq:def-Q}
\end{align}
Here, $\Tilde{I}$ is the label of the vector charged object when inserted multiple times, and $p$ is the label of plaquettes in $M_2-\sum_{\Tilde{I}}\mathcal{D}_{\Tilde{I}}$. From this relation~\eqref{eq:vector-selection}, we can show that the vector charge is saturated by the first Chern number $Q$, as expected from the index theorem.

According to the bosonization rule, the axial charged operator is a vertex operator, and it is generally defined as
\begin{align}
     V_{\{n_\alpha \}}(n):=e^{i\sum_\alpha n_\alpha \phi_\alpha(n)}.\label{eq:def-vertex-op}
\end{align}
Since $V_{\{n_\alpha \}}(n)
   \to\exp\left[-i\sum_\alpha 2q_{A,\alpha}n_\alpha \Lambda(n)\right]
   V_{\{n_\alpha \}}(n)$ undergoes a gauge transformation, we need to attach an open Wilson line to make it gauge-invariant:
   \begin{align}
        V_{\{n_\alpha \}}(n) \exp\left[i\sum_\alpha 2q_{A,\alpha}n_\alpha\sum_{(n,\mu)\in P}^n A_\mu(n)\right].
        \label{eq:gauge-inv-vertex-operator}
   \end{align}
   Here, $P$ is the path with the endpoint $n$. Let us now consider the condition for the correlation function to be non-zero when Eq.~\eqref{eq:gauge-inv-vertex-operator} is inserted. For this purpose, we require that the correlation function is invariant under a constant shift $\phi_\alpha\to\phi_\alpha+\xi_\alpha$. Hence, we obtain the following selection rule:
   \begin{align}
       \sum_I n_{I,\alpha}
   &=\frac{q_{V,\alpha}}{2\pi}\sum_{\Tilde{p}\in\Tilde{M}_2} F_{12}(\Tilde{p})\label{eq:axial-selection}
   =q_{V,\alpha}\Tilde{Q}\\
   \Tilde{Q}&:=\frac{1}{2\pi}\sum_{\Tilde{p}\in\Tilde{M}_2} F_{12}(\Tilde{p})\label{eq:def-tilde-Q}
   \end{align}
   Here, $I$ is the label of the vector charged objects, and $\Tilde{p}$ is the label of the dual plaquettes in $\Tilde{M}_2$.

Now, combining Eqs.~\eqref{eq:vector-selection} and \eqref{eq:axial-selection}, we would like to consider the selection rule for left-handed and right-handed fermions. However, there are two Chern numbers, $Q$ and $\Tilde{Q}$, which are generally not equal. However, since $Q - \Tilde{Q} = (\text{Finite sum of }\frac{F}{2\pi}\text{ near }\mathcal{D}) \in \mathbb{Z}$, and $|Q - \Tilde{Q}|$ can be bounded by using $\delta$ and $\delta'$. Hence, assuming sufficiently strict admissibility, we can justify $Q = \Tilde{Q}$.

From the gauge transformation properties and their relative signs of $V_{\{n_\alpha \}}(n)$ and $M_{\{m_\alpha\}}(\mathcal{D})$, the following relations can be deduced:
\begin{align}
&P_R \psi_\alpha : e^{+i\phi_\alpha(n)/2} M_{m_\alpha=-1}(\mathcal{D}), \quad \Bar{\psi}_\alpha P_L : e^{-i\phi_\alpha(n)/2} M_{m_\alpha=+1}(\mathcal{D}).
\notag\\
   &P_L \psi_\alpha : e^{-i\phi_\alpha(n)/2} M_{m_\alpha=-1}(\mathcal{D}), \quad \Bar{\psi}_\alpha P_R : e^{+i\phi_\alpha(n)/2} M_{m_\alpha=+1}(\mathcal{D}).\label{eq:RL}
\end{align}
Also, when $Q = \Tilde{Q}$, we have
\begin{align}
    \sum_I n_{I,\alpha} + \frac{1}{2} \sum_{\Tilde{I}} m_{\Tilde{I},\alpha} = q_{L,\alpha} Q, \quad
    \sum_I n_{I,\alpha} - \frac{1}{2} \sum_{\Tilde{I}} m_{\Tilde{I},\alpha} = q_{R,\alpha} Q.\label{eq:combined}
\end{align}
From this, we can see that, for example, $P_R \psi_\alpha$ contributes to $q_{L,\alpha}Q > 0$, while $P_L \psi_\alpha$ contributes to $q_{R,\alpha}Q < 0$. In summary, we can obtain selection rules that are consistent with the fermion number anomaly in the continuum theory (see Ref.~\cite{Fujikawa:1994np} and references cited therein.)
\begin{align}
     \partial_\mu J_\mu^{L,R}(x) = \mp \frac{q_{L,R}}{2\pi} F_{12}(x).
     \label{fermion-number-anomaly}
\end{align}

\section{Conclusion}
We have achieved a lattice regularization of 2D $U(1)$ chiral gauge theory based on bosonization, using a Wilson-type lattice regularization with compact variables. In this formulation, we realized the expected topological nature from the continuum theory by imposing the admissibility condition on lattice fields, and represented vector-charged objects as ``holes'' excised from the lattice to ensure consistency with the admissibility condition. Furthermore, we have shown that our lattice definitions of charged objects reproduce selection rules consistent with the fermion number anomaly in the continuum theory.

In addition, we note a possible future direction for the present work. According to non-Abelian bosonization, the fundamental variable is the $U(N)$-valued compact variable. Furthermore, for the topological nature of the Wess--Zumino--Witten term on the lattice, sufficient smoothness of the lattice fields plays a crucial role (see, for example, the construction in Ref.~\cite{Shiozaki:2024yrm}). Therefore, our work may play a significant role
in the context of constructing non-Abelian chiral gauge theories
based on non-Abelian bosonization~\cite{Witten:1983ar}.\footnote{%
One may not respect the admissibility condition
for bosonized Abelian lattice gauge theories.
For instance, Ref.~\cite{Berkowitz:2023pnz} realizes vector symmetry
and its charged objects
by explicitly introducing a dual scalar field~$\Tilde{\phi}$
based on the modified Villain formulation.
Non-Abelian Villain formulation is also an intriguing issue~\cite{Chen:2024ddr}.}

\subsection*{Acknowledgements}
This work was partially supported by
Kyushu University’s Innovator Fellowship Program (S.O.)
and Japan Society for the Promotion of Science (JSPS)
Grant-in-Aid for Scientific Research Grant Numbers
JP22KJ2096 (O.M.) and JP23K03418 (H.S.).
O.M.\ acknowledges the RIKEN Special Postdoctoral Researcher Program.

\bibliographystyle{JHEP}
\bibliography{ref}

\end{document}